\documentclass{emulateapj}
\usepackage{apjfonts}
\usepackage{amsmath}

\newcommand{\msun}{M$_{\sun}$}
\newcommand{\kms}{km s$^{-1}$}
\newcommand{\msuns}{M$_{\sun}~$}
\newcommand{\kmss}{km s$^{-1}~$}
\newcommand{\msunyr}{M$_{\sun}$ yr$^{-1}~$}

\makeatletter
\def\fnum@figure{{\footnotesize\bf Figure\space\thefigure.~}}
\def\fnum@table{{\bf Table~\thetable}}
\makeatother

\begin{document}

\title{BONDI-HOYLE-LYTTLETON ACCRETION ONTO A PROTOPLANETARY DISK}

\author{Nickolas Moeckel\altaffilmark{1}}
\affil{SUPA, School of Physics and Astronomy, University of St Andrews, North Haugh, St Andrews, Fife, KY16 9SS \\and School of Physics, University of Exeter, Stocker Road, Exeter EX4 4QL}
\author{Henry B. Throop}
\affil{Southwest Research Institute, Department of Space Studies, 1050 Walnut St, Ste 300, Boulder, CO 80302, USA}

\email{nickolas1@gmail.com}

\begin{abstract}
Young stellar systems orbiting in the potential of their birth cluster can accrete from the dense molecular interstellar medium during the period between the star's birth and the dispersal of the cluster's gas.  Over this time, which may span several Myr, the amount of material accreted can rival the amount in the initial protoplanetary disk; the potential importance of this `tail-end' accretion for planet formation was recently highlighted by \citet{throop08}.  While accretion onto a point mass is successfully modeled by the classical Bondi-Hoyle-Lyttleton solutions, the more complicated case of accretion onto a star-disk system defies analytic solution.  In this paper we investigate via direct hydrodynamic simulations the accretion of dense interstellar material onto a star with an associated gaseous protoplanetary disk.  We discuss the changes to the structure of the accretion flow caused by the disk, and {\em vice versa}.  We find that immersion in a dense accretion flow can redistribute disk material such that outer disk migrates inwards, increasing the inner disk surface density and reducing the outer radius.  The accretion flow also triggers the development of spiral density features, and changes to the disk inclination.  The mean accretion rate onto the star remains roughly the same with and without the presence of a disk.  We discuss the potential impact of this process on planet formation, including the possibility of triggered gravitational instability; inclination differences between the disk and the star; and the appearance of spiral structure in a gravitationally stable system.
\end{abstract}


\section{Introduction}
Generally, stars form in clusters \citep[e.g.][]{lada03}, and our current understanding of star formation is beginning to incorporate the effects of the cluster environment on the process rather than treating star formation as a one-system event occurring in isolation.  A complete picture of planet formation may require a similar awareness of the surroundings; the possible influence of the cluster on planet formation is a logical consequence of clustered star formation.  Protoplanetary disks have lifetimes of $\sim 1$--10 Myr without significant external UV irradiation \citep[e.g.][]{currie07,currie09}.  In the Orion Nebula Cluster, 80\% of stars have disks \citep{smith05}, and even in environments like the Trapezium cluster over 10\% of stars may have disks containing a minimum mass solar nebula (MMSN, $\sim 0.01$ \msun) of material \citep{mann09}.  These disk lifetimes are comparable to or longer than the crossing time of a typical star-forming region, and thus the cluster environment may affect planet formation or young planetary systems. 

Disruptive possibilities include the truncation of disks  by interactions with other cluster members \citep{clarke93,heller95,boffin98} or dynamically altering an existing planetary system \citep{laughlin98,zakamska04,malmberg09}\footnote{Although the rates for these interactions are probably low compared to formation timescales \citep{adams06}.}, and photo-evaporation which can remove disks on $10^5$--$10^7$ yr timescales \citep{hollenbach94,johnstone98}.  More constructive effects may include triggering planetesimal formation by external UV illumination \citep{throop05}, and adding material to the disk by accretion from the cluster's remnant ISM reservoir \citep[][hereafter TB08]{throop08}

The effect that we explore in this work is the accretion of ISM material by a star--disk system.  Bondi--Hoyle--Lyttleton accretion by a star moving through a background medium may play a fundamental role in determining the final stellar mass \citep{bonnell97, bonnell01a}, an idea that has seen a large amount of study over the last decade.  A less-explored question is the importance of accretion after a star has already assembled the majority of its mass, and the cluster is dominated by the gravitational dynamics of the stars rather than the potential and bulk velocities of the molecular gas.  

TB08 performed simulations of this `tail-end' accretion, finding that as stars orbit through the depleting molecular gas in a cluster they can accrete around 1 MMSN per 1 Myr across a range of standard star-forming regions.  The observational constraints on disk lifetimes allow for this accretion to take place while a primordial protoplanetary disk is still present.  A plausible example of the interaction between a disk and the ISM is found around the star HD 61005 \citep{hines07}, displaying what appears to be a wind-swept disk.  The TB08 study highlighted the potential importance of this process and suggested several possible measurable effects,
including altering the structure of the protoplanetary disk, changing its orientation, or changing its composition.

It is these ideas that we explore here via direct hydrodynamic simulations. TB08 treated accretion by an analytic prescription applied post-simulation to {\it n}-body experiments, and here we present the results of exploratory simulations involving a protoplanetary disk that encounters a dense ISM at supersonic speeds.  In TB08 the accretion is treated as occurring from a mean background density over $\sim 10^6$ yr; the variation we consider is a star-disk system encountering a dense filament or clump, with a higher accretion rate over a shorter amount of time.  After describing our setup and presenting the results of our experiments, we discuss their implications for theories of planet formation.

\section{Physical and Numerical Model}
\subsection{Bondi-Hoyle-Lyttleton Accretion}
Accretion onto a point mass moving with some velocity through an ambient medium is a well-studied question, and the classical solutions developed by Bondi, Hoyle and Lyttleton (BHL) \citep{hoyle39, bondi44, bondi52} are widely used as a starting point (and often as an end point) when considering accretion problems.  Many numerical tests have added to our understanding, offering verification and modification to the analytic accretion rates \citep{shima85,ruffert94,ruffert94a,ruffert94b,ruffert95,ruffert96,ruffert99} and considering complications like the effects of radiation \citep{edgar04a}, vorticity \citep{krumholz05}, and turbulence \citep{krumholz06}.  A good review of the historical development and numerical testing of BHL accretion theory is found in \citet{edgar04}.

The general picture of BHL accretion is of a point mass in motion through a uniform, infinite medium.  Ambient material is focused into a wake or accretion column behind the accretor, where colliding streamlines cancel out velocities that aren't parallel to the body's motion.  Material passing within some critical radius becomes bound and is energetically doomed to flow up the wake and be accreted by the star.  Given the density and sound speed of the ambient medium, and the velocity and mass of the accretor, the critical radius can be calculated and thus the accretion rate.  The analytic expression for the accretion rate varies by a few factors of 2 in the literature, and we briefly explain our choices here.

The critical impact parameter of a test particle moving at velocity $v$ past a body of mass $M$ can be found by simply equating the potential and kinetic energies, and one finds the Hoyle--Lyttleton radius 
 $ R_{HL} = 2 G M/v^2$ \citep{hoyle39}.
For highly supersonic motion, accretion is firmly in the Hoyle--Lyttleton limit and this would be a fine approximation to the accretion radius for this work.  For ease of comparison to recent numerical results, we will use as our accretion radius the interpolation between the supersonic and stationary accretion regimes proposed by \citet{bondi52} with the factor of 2 modification suggested by \citet{shima85},
\begin{equation}
  R_{A} = \frac{2 G M}{c_s^2 + v^2},
  \label{accrad}
\end{equation}
and an associated accretion rate
\begin{equation}
  \dot{M}_{BH} = \pi R_{A}^2 \rho (\lambda^2 c_s^2 + v^2)^{1/2}.
  \label{mdotbh}
\end{equation}
The constant $\lambda$ is of order unity and depends on the equation of state of the gas; for the isothermal limit, $\lambda = e^{3/2}/4 \sim 1.120$ \citep{bondi52}.  The mass density and sound speed of the ambient medium are $\rho$ and $c_s$, respectively.

\subsection{Parameters and Initial Conditions}
Throughout this work, we take the mass of the accretor to be $M = 1$ \msun, and its velocity through the medium $v = 3$ \kms.  We assume the medium has a mean molecular weight $\mu = 2.3$ and a sound speed $c_s = 0.3$ \kms,  thus a  temperature $\sim 25$ K and a Mach number of 10.  The number density of the medium is set at $n = 5 \times 10^6$ cm$^{-3}$, yielding a mass density $\rho = 1.92 \times 10^{-17}$ g cm$^{-3}$.  Plugging these values into equations \ref{accrad} and \ref{mdotbh} gives us our reference accretion radius and rate and the natural timescale for the problem,
\begin{align}
R_A & = 195 ~{\rm AU} \nonumber \\
\dot{M}_{BH} & =  2.47 \times 10^{-6} ~{\rm M}_{\sun} ~{\rm yr}^{-1} \nonumber \\
\tau_{BH} & = R_A/c_s \approx 3000  ~{\rm yr}.
\label{referencerates}
\end{align}
These ISM density values are quite high, and are more associated with the regions surrounding dense cores \citep[which can reach mean hydrogen densities in excess of $10^7$ cm$^{-3}$ at 0.05 pc from the core, e.g.][]{mezger90} or over-dense ridges \citep[e.g. a 0.5 pc ridge with $n>10^6$ cm$^{-3}$ near the BN/KL region,][]{kawamura02}, rather than the ambient background.  Thus our simulations are only directly applicable in the densest regions of molecular clouds.  This ISM density is picked for computational reasons discussed below, rather than purely physical motivation.
 
Our simulations model a gaseous protoplanetary disk immersed in the ISM flow.  
The disk initially contains 0.01 \msun, extending from 10 - 100 AU with a surface density profile $\Sigma \propto r^{-3/2}$.  For reference, this is comparable to the density of a classical MMSN \citep{weidenschilling77}, and would give a surface density $\sim 1000$ gm cm$^{-2}$ at 1 AU were the disk to extend in that far.  The disk and the inflowing material are both isothermal at the same temperature.

We carried out four simulations: accretion from a disk with no ambient medium; accretion onto a point mass (BHL accretion); BHL accretion with a disk perpendicular to the inflow; and BHL accretion with the disk inclined $30^\circ$ to the inflow.  The first two experiments serve to set baseline accretion rates onto the star from the ISM and the disk in isolation, while the experiments with both inflow and a disk explore how the disk's presence alters the ISM accretion flow and {\it vice versa}.  We run each simulation until $t \approx 5000$ yr, during which the star or star-disk is immersed in the accretion flow for 1 $\tau_{BH}$ and evolves for a further 0.5 $\tau_{BH}$ after exiting the ISM.  This is sufficient time for a quasi-stable accretion state to be set up.  We will refer to time in years as $t$, and time in units of $\tau_{BH}$ from as $\tau$.  The zero point for each timescale is set to when the ISM flow first reaches the star.  Each system evolves in effective isolation for $t \sim 525$ yr leading up to $t=0$, which allows the disks to equilibrate and gives the ISM time to traverse from the inflow boundary to the star's position.  The inflow boundary is turned off such that after being immersed in the flow for 1 $\tau_{BH}$ the star exits the dense medium, the remaining gas is accreted or leaves the domain, and the disk settles down.  We can then see the effect on a protoplanetary disk of a passage through a dense clump or filament in a molecular cloud.

In TB08, the motivating work for this study, the stars orbited and accreted over a period of 4 Myr, accreting from a Plummer sphere gas distribution with a peak density ranging from $10^4$--$10^6$ cm$^{-3}$ depending on the total cluster mass.  Our background density is five times higher than the nominal peak density\footnote{A Plummer sphere with 3000 \msuns of gas matching Orion's current core radius of $\sim 0.2$ pc \citep{hillenbrand98} has a central density $\sim 10^6$ cm$^{-3}$.} in an Orion molecular cloud analog, $10^6$ cm$^{-3}$, thus our picture of a collision with a prestellar core or filament.  This choice is motivated by computational necessity rather than just physical arguments, as such collisions will be very rare in all but the densest protoclusters.  A simple estimate\footnote{For example, taking the $m =1$ \msuns protostars or cores to have a collision radius $r=0.01$ pc, number density $n_\star=10^4$ pc$^{-3}$, and velocity dispersion $\sigma = 2$\kms (appropriate for the Trapezium), the standard collision time \citep{binney08} is $t_{coll}^{-1} = 4 \sqrt{\pi} n_\star \sigma (r^2 + G m r / \sigma^2) = 6.6\times10^5$ yr.  The scaling is linear with number density, so a cluster with $n_\star=10^3$ would have an encounter rate in excess of 6 Myr.}
of the encounter rate between a protostar and a collapsing core yields encounter times of at least $\sim$ 0.7 Myr.  While this is short enough to allow for a handful of such encounters over the cluster lifetime assuming continuing star formation for a few Myr, it is longer than the likely collapse time for dense prestellar cores \citep[a few $\times 10^4$--$10^5$ yr, e.g.][]{ward-thompson07}, which is the more relevant time constraint.  As described more fully in section \ref{thecode}, we have chosen to minimize the density disparity between the disk and ISM, pushing the disk mass to low values and the ISM density to unrealistically high values.  Additionally, we simulate the interaction only long enough to set up a quasi-stable accretion flow before ending the simulations, so that the disk is left as unaffected as possible by the numerical viscosity inherent in any computational scheme.  These are the constraints that motivate our scenario of an encounter with a dense clump or core rather than the background ISM as in TB08.  The time-scale arguments above suggest that these specific simulation parameters are directly applicable to only a few percent of stars in an Orion-like environment.  However, as we argue in the next section, the main effect of the ISM accretion on the disk is not impulsive but rather a function of the total mass accreted, and so the results of these over-dense simulations should be broadly applicable to the original scenario in TB08.

\subsection{Expected Effects}
\label{estimations}
There are several effects that one can imagine occurring when a gas disk is subjected to what is effectively a supersonic wind, including modifications to the disk structure and the morphology of the accretion column.  Here we discuss a few of them and estimate how likely they are to play a role given our chosen parameters.

{\em Erosion by ram pressure stripping:} Depending on the parameters of the ISM, disk density, stellar mass and velocity, the wind may truncate the disk at some radius.  Ram pressure stripping is frequently considered in studies of galaxies interacting with the intra cluster medium, a situation for which several analytic estimates of the truncation radius have been developed \citep{gunn72,mori00}.  More relevant to this work, the question of supernovae shells interacting with protostellar disks was discussed by \citet{chevalier00}.  As in these works, we compare the ram pressure of the ISM flow to the gravitational restoring force keeping the disk associated with the star.  The radius at which these are comparable should give some indication of the importance of ram pressure stripping to our 100 AU disk, which we assume here to be face-on to the ISM flow.

The gravitational force per unit area (the gravitational `pressure' to which we can compare the ISM ram pressure) on a ring of radius $r$ is given by $P_{grav} = G M \Sigma(r) / r^2$.  For our disk with $\Sigma(r) \approx 1000(r /{\rm AU})^{-3/2}$ g cm$^{-2}$, this is $P_{grav} \approx 600 (r/{\rm AU})^{-7/2}$ dyne cm$^{-2}$.
The ram pressure of the ISM flow is given by $P_{ISM} = \rho v^2 \approx 1.7 \times 10^{-6}$ dyne cm$^{-2}$.  The radius at which $P_{ISM} > P_{grav}$ is approximately 275 AU, larger than the initial disk radius of 100 AU.    Bearing in mind the order-of-magnitude nature of this calculation, especially when estimating $P_{grav}$, it would not be surprising to see ram pressure affecting the extremities of the disk, but it seems clear that its effect should be negligible for the inner regions of the disk.  Note that outside of 75 AU the gas pressure in the disk $P_{disk} = c_s^2 \rho \lesssim 1.7\times10^{-7}$ dyne cm$^{-2}$, one order of magnitude less than $P_{ISM}$.  Flattening, but not unbinding, of the disk at the outer edge is to be expected.

{\em Momentum stripping:} The ram pressure stripping calculation assumes that a  quasi-equilibrium exists between the pressure of the ISM wind and the gravity in the star-disk system, and thus that nothing changes on roughly the orbital timescale of the disk.  A potentially shorter-timescale effect (shorter than, say, the the orbital period at the outer disk radius) is the addition of momentum perpendicular to the disk rotation.  If the momentum addition raises the disk material to escape velocities, the disk will be eroded from the outer edge \citep{chevalier00}.

The escape velocity of disk material at radius $r$ is $v_{esc} = (2 G M / r)^{1/2}$.  Since the material is moving at the Keplerian velocity already, assuming again a face-on disk we require a velocity gain perpendicular to the disk plane $\delta v_z(r) = (G M / r)^{1/2}$ on a timescale shorter than the orbital period to unbind disk material at radius $r$.  The maximal change in velocity would occur if the disk swept up much more than its own mass (which it won't with our parameters), in which case the disk material would move with the ISM velocity of 3 \kms.  This is less than the required velocity change even at 100 AU, so the disk is very robust to erosion by absorbing momentum from the ISM.

{\em Disk redistribution:}
While absorbing momentum from the ISM will not unbind our disk at any radius, it may have an effect on the radial distribution of material.
The disk mass associated with the annulus of infinitesimal radius $dr$ at radius $r$ is $m_{disk} = 2 \pi r \Sigma(r) dr$, while the swept-up ISM material after some time $t$ is $m_{ISM} = 2 \pi r \rho v t dr$. 
When the ISM material is absorbed, the specific angular momentum in that annulus drops by a factor of $\Sigma(r)/[\Sigma(r) + \rho v t]$.  Assuming that the annulus adjusts to the radius whose Keplerian angular momentum reflects this new value, the new radius $\tilde{r}$ can be found from $\tilde{r}^{1/2} \approx r^{1/2} \Sigma(r)/[\Sigma(r) + \rho v t]$. 

 This rough calculation ignores any interaction with neighboring annuli, and assumes a smooth addition of material to the disk.  Those caveats in mind, after accreting for a time $t = \tau_{BH}$, material at 50 AU will have migrated to 48 AU, while the outer edge of the disk will have moved from 100 AU to 90 AU.  The trend of this effect is for the disk to become more compact, with the outer edges migrating proportionally further in.

Summarizing the likely effects of the ISM flow on the disk: ram pressure will only noticeably alter the outskirts of the disk.  Mass and momentum loading will not unbind the disk in any way, but will redistribute material inwards, and this effect will increase with radius.  We see that even at the very high ISM densities we are simulating, the ram pressure is not the dominant effect.  Rather, the total mass that is absorbed (over timescales longer than the orbital period of the disk) dominates the disk evolution.  At more realistic ISM densities the ram pressure effects, already less important than mass loading, will be further diminished.  The total mass accreted by our systems over their 3000 yr immersion in the dense ISM, roughly $\dot{M}_{BH}t \sim 7.5\times10^{-3}$ \msun, is less than that accreted by the average solar-mass star in TB08, and thus the disk redistribution should be comparable to that seen here.

{\em Alteration of the accretion flow:} We estimated above the unimportance of the ISM ram pressure in influencing the disk evolution; conversely, this suggests that the disk will alter the accretion flow compared to the diskless case.  Since the disk radius is $r_{disk} \approx R_A / 2$ and is centered on the axis of the accretion flow, one-quarter of the accretion flow is intercepted by the disk.  The intercepted material is on streamlines that focus to the accretion axis close in to the star.  Without these converging streamlines compressing the wake, the accretion structure close to the star should be less collimated in the presence of a disk, as the final inflow occurs in the `lee' of the disk.

\subsection{The Code}
\label{thecode}
We used the smoothed particle hydrodynamics (SPH) code GADGET-2 \citep{springel05} to perform our 3D hydrodynamic simulations.  The code has been modified to suit this problem in the following ways:  

{\em Sink particles:} The accretor is treated as a sink particle \citep{bate95}, accreting all particles that pass within its accretion radius and absorbing their linear momentum.  By removing particles that would otherwise have a tiny timestep we are able to follow the evolution of the system for a longer time.  We apply no boundary conditions at the accretion radius, which we set at $R_{acc}= 9.75 ~{\rm AU} = 0.05 ~R_A$; this value is a compromise between computational efficiency and the reality of a nearly-point mass accretor.  The comprehensive series of BHL accretion simulations by Ruffert showed that with accretors larger than this, the character of the flow is fundamentally changed as accretion is onto the leading edge of the body rather than along the accretion column.

{\em Artificial viscosity:} SPH codes rely on an artificial viscosity to capture shocks and prevent particle interpenetration.  In place of the GADGET-2 default viscosity, we have implemented the viscosity suggested by \citet{morris97}.  Here each particle has its own viscous parameter $\alpha$, which evolves according to
\begin{displaymath}
    \dot{\alpha} = -(\alpha - 0.1)/\tau_\alpha + {\rm max}(-|{\bf \nabla} \cdot {\bf v}|,0)(2 - \alpha).
\end{displaymath}
  
The first term decays the viscosity down to the minimum value 0.1 over the timescale $\tau_\alpha$, while the second term is meant to detect shocks and cause the viscosity to grow.  Each particle's individual $\alpha$ then figures in the standard viscosity formulation, rather than a fixed value for all particles.  This reduction in the numerical viscosity away from shocks gives a less dissipative treatment of smooth flows and quiescently rotating disks.

{\em Inflowing and outflowing boundary conditions:} A critical element of our study is the inflow of ambient ISM material into the simulation domain.  We have implemented boundary conditions such that the star sits in a cylinder with height and diameter 5 $R_A$.  The star is placed on the cylinder's axis, 2.5 $R_A$ from both the inflow and outflow boundaries.  The ISM flow is introduced at the inflow boundary and consists of randomly placed SPH particles at the chosen velocity and mass.  A boundary layer of particles that experience no acceleration shepherds the ISM flow at the outer edge of the cylinder.  SPH particles that leave the domain through the outflow edge are removed from the calculation.  These are not periodic boundary conditions; the ISM material introduced at the inflow boundary is `fresh'.

This calculation is challenging in that there is a large mismatch between the densities of the ISM and the disk, which we have ameliorated as much as possible by choosing a dense ISM and a low-mass disk.  SPH codes use a Lagrangian scheme in which the fluid is represented by discrete tracer particles, which are interpolated over a density-dependent smoothing length to obtain continuum values throughout the domain.  Choosing equal mass ISM and disk particles yields an undesirably large mismatch in smoothing lengths, and setting the masses such that the smoothing lengths are equal gives unacceptably disparate masses.  By setting the disk particle mass to be 8 times the ISM particle mass we avoid the problems associated with large particle mass ratios, while still having the ISM particle smoothing lengths set such that the ISM flow comfortably resolves the protoplanetary disk.  The disk is initially modeled with $\sim 2.5 \times 10^5$ particles; the total particle number is constantly varying, but when the cylinder is full there are $\sim 5 \times 10^6$ ISM particles.

\section{Numerical Results}
\begin{figure}
\centering
  \plotone{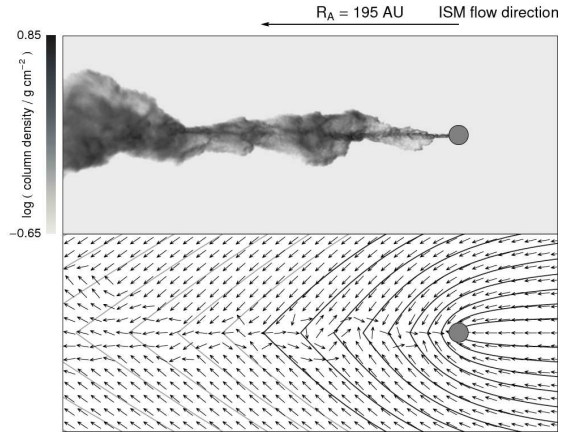}
  \caption{{\it Top:} column density plot of Mach 10 BHL accretion, showing only material 2.5 times denser than the background density.  The ISM flows in from the right and exits left.  The accretor is shown as a gray circle.  {\it Bottom:} black arrows indicate the velocity field of the flow, showing direction only.  Solid grayscale streamlines are the analytic flow field of the Hoyle-Lyttleton solution.  {\it Dark gray} streamlines are interior to the critical impact parameter, while unbound streamlines are in {\it light gray}.  The stagnation point of the simulated accretion column matches the analytic value, though the streamlines differ somewhat near the accretion axis.}
  \label{nodiskdensvel}
\end{figure}

\begin{figure*}
\centering
  \plotone{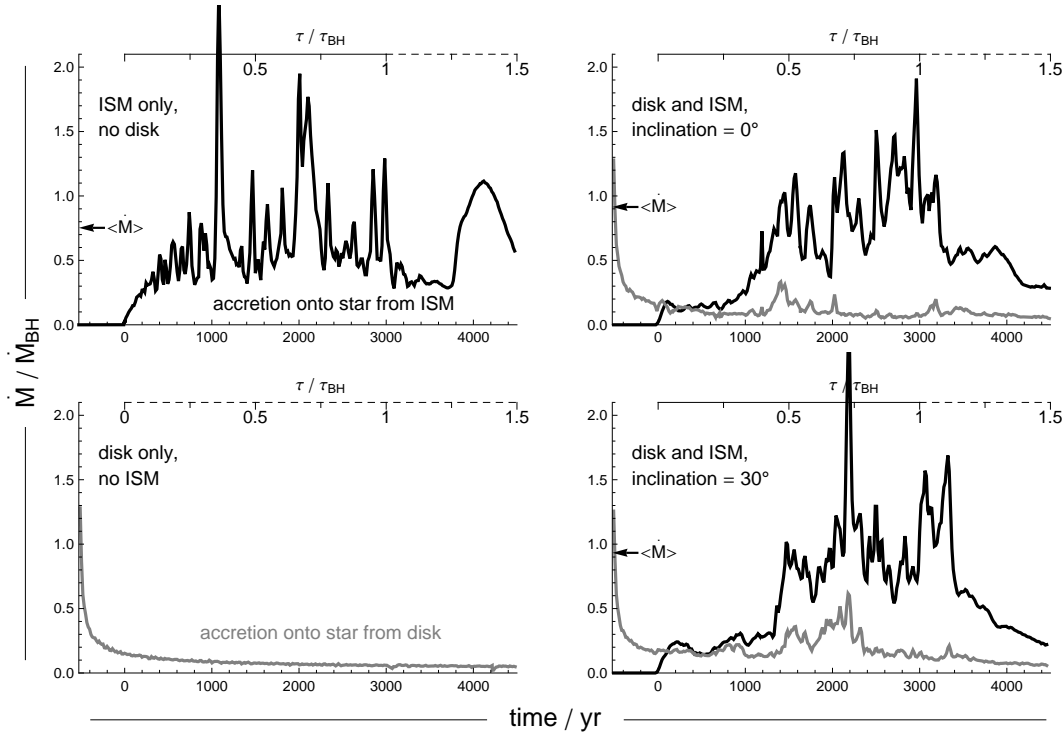}
  \caption{Accretion rates onto the star from the two sources in the simulations; {\it black} lines are accretion from the ISM, while {\it gray} lines are accretion from the disk.  Arrows at the left of each panel show the mean ISM accretion rate for times $0.5 \leq \tau / \tau_{BH} \leq 1.0$.  The time scale at the top of the panels is solid when the ISM flow is on, and dashed when the star has left the flow; the star enters the dense ISM at at $\tau = 0$, and remains in it until $\tau = \tau_{BH}$, and we run the simulations for an additional $0.5\tau_{BH}$ after the end of the ISM flow has passed by the star.  {\it Top left:} accretion from the ISM with no disk.  {\it Bottom left:} baseline accretion rate from the disk with no ISM.  {\it Top right:} both accretion rates for the case with the disk face-on to the inflow.  {\it Bottom right:} both accretion rates with the disk inclined 30$^\circ$ to the inflow.}
  \label{accretionrates}
\end{figure*}

\subsection{BHL Accretion With No Disk}
We begin by examining the morphology of the accretion flow with no disk present.  Our isothermal inflow is Mach 10 and our accretor is 0.05 $R_A$ in radius, a setup which lies between models HS and HM in \citet{ruffert96}, to which we can compare our results.  Figure \ref{nodiskdensvel} shows the column density\footnote{We used the program SPLASH \citep{price07} to project the SPH particle positions into column density values and velocity fields for this and similar plots.} and velocity field for this run once the accretion flow is established.  In this, and all surface density plots in the paper, we plot only the contribution to the column density from particles with density greater than 2.5 times the background.  This effectively isolates the shocked accretion column from the surrounding ISM, removing the arbitrary scaling associated with the size of the accretion domain as well as removing some artifacts from the boundary layer of particles.  The simulation domain is larger than the region shown; we show here just the accretion column within 2 $R_A$ of the star, while the simulation extends 2.5 $R_A$ in radius, upstream, and downstream from the star.

The column density plot appears largely as one would expect.  With high Mach number flows the accretion column has a small opening angle, tightly compressed behind the accretor and widening as the distance from the accretor increases.  We do see larger breaks from axisymmetry in our run than Ruffert did, an effect that also manifests itself in our accretion rates.  A possible cause of this is the way we introduce the accretion flow, randomly seeded in space rather than on a fixed grid.  The random placement of our SPH particles, while maintaining an overall mass inflow rate that matches our chosen value, allows for random pockets of over- and under-density that do not smooth themselves out before reaching the accretion shock.  These may act to seed instability in the accretion column.

The velocity field shows the fluid direction only, not the magnitude, overlaid on the analytic streamlines of the Hoyle-Lyttleton solution from \citet{bisnovatyi-kogan79}.  The darker streamlines are expected to remain bound to the accretor, while lighter streamlines should escape.  The position of the stagnation point in our simulation is variable in time, but always remains in the vicinity of the expected location.  While the character of the ballistic streamlines matches the simulation, the detailed shape is different, especially as the flow approaches the accretion axis.

In figure \ref{accretionrates} we show the accretion rates for all four simulations.  The upper left panel shows the BHL accretion case.  There is no accretion until the inflow turns on and reaches the star at $\tau = 0$.  The accretion rate ramps up to an oscillating, quasi-steady state at $\tau \sim 0.5~\tau_{BH}$.  The large spike in the accretion rate at $\tau \sim 0.3~\tau_{BH}$ is from when a clump that formed as the edge of the accretion flow converged downstream, stagnated, and finally accreted back onto the star.  We believe this to be an artifact of the way the star effectively enters the dense region abruptly from a zero-density region, rather than beginning immersed in the ambient medium.  The increase in the accretion rate at $\tau \sim 1.25~\tau_{BH}$ is a result of the star leaving the dense ISM.  Gas velocities near the stagnation point are subsonic, and near the end of the accretion period a large blob of gas forms and drifts slowly away from the star and out of the domain.  When the dense ISM ends and there is no pressure from the upstream direction, this gas flows back toward the star and is accreting as the simulation ends.  We measure the mean accretion rate onto the star between $0.5 \leq \tau / \tau_{BH} \leq 1$, i.e. between when the accretion column is fully established and before the stars leaves the dense ISM, as $<\dot{M}> = 0.77 \dot{M}_{BH}$.  This agrees within $\sim 6$\% of the roughly equivalent case in \citet{ruffert96} of $<\dot{M}> = 0.82 \dot{M}_{BH}$, though we have more variation about the mean than in his simulations.

\subsection{Disk Evolution With No Inflow}
In order to get a baseline accretion rate from the disk with no ISM accretion, we ran the disk in isolation for the same amount of time.  The accretion rate is shown in the lower left panel of figure \ref{accretionrates}.  The accretion rate is marked by a high initial value as the inner edge of the disk drains onto the star, before settling down to a low value of $\sim 1.5 \times 10^{-7}$ \msunyr.

\begin{figure}
\centering
  \plotone{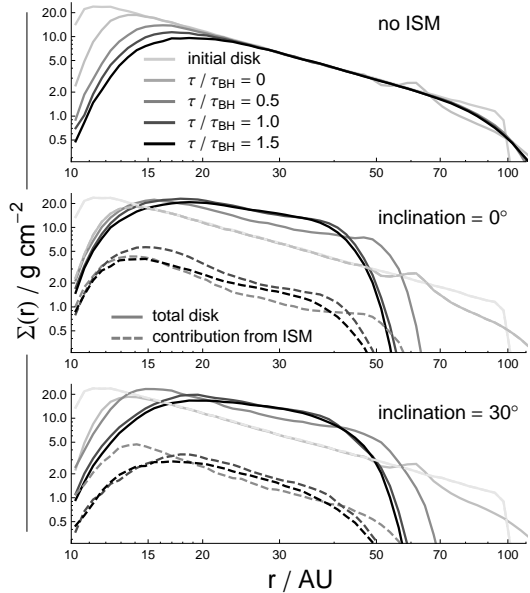}
  \caption{{\it Top:} azimuthally-averaged surface density profile for the disk evolving in isolation.  {\it Middle:} evolution of the disk face-on to the inflow.  The gray levels are as in the top panel.  Solid lines are the total surface density, and dashed lines are the contribution to the disk from the ISM.  {\it Bottom:} lines are as for the middle panel, but for the case with the disk inclined 30$^\circ$ to the inflow.}
  \label{surfacedensities}
\end{figure}

The top panel of figure \ref{surfacedensities} shows the evolution of the azimuthally-averaged surface density profile for the isolated disk case.  The lightest line shows the profile just after the simulation began, and further lines are at values of $\tau_{BH}/2$ starting at $\tau = 0$ (recall that this is the time when the accretion flow encounters the disk).  The wiggle in the density profile at $\tau = 0$ is the final transient of the disk equilibration.  Throughout the run the inner edge of the disk drains onto the accretor.  This is due predominantly to the numerical dissipation from the artificial viscosity, with a small contribution from the lack of gas pressure support from SPH particles starward of the inner edge.  Boundary conditions at the accretor might prevent this drainage, but rather than risk introducing other effects we allow the disk to evolve under the influence of artificial viscosity.  There is some spreading at the outer edge as well, but the density profile over the bulk of the disk remains almost unchanged.  Throughout the simulation the disk is stable to gravitational instabilities, and no spiral features develop.

\subsection{BHL Accretion With Disk Inclined $0^{\circ}$ To Inflow}
\begin{figure*}
\centering
  \epsscale{0.75}
  \plotone{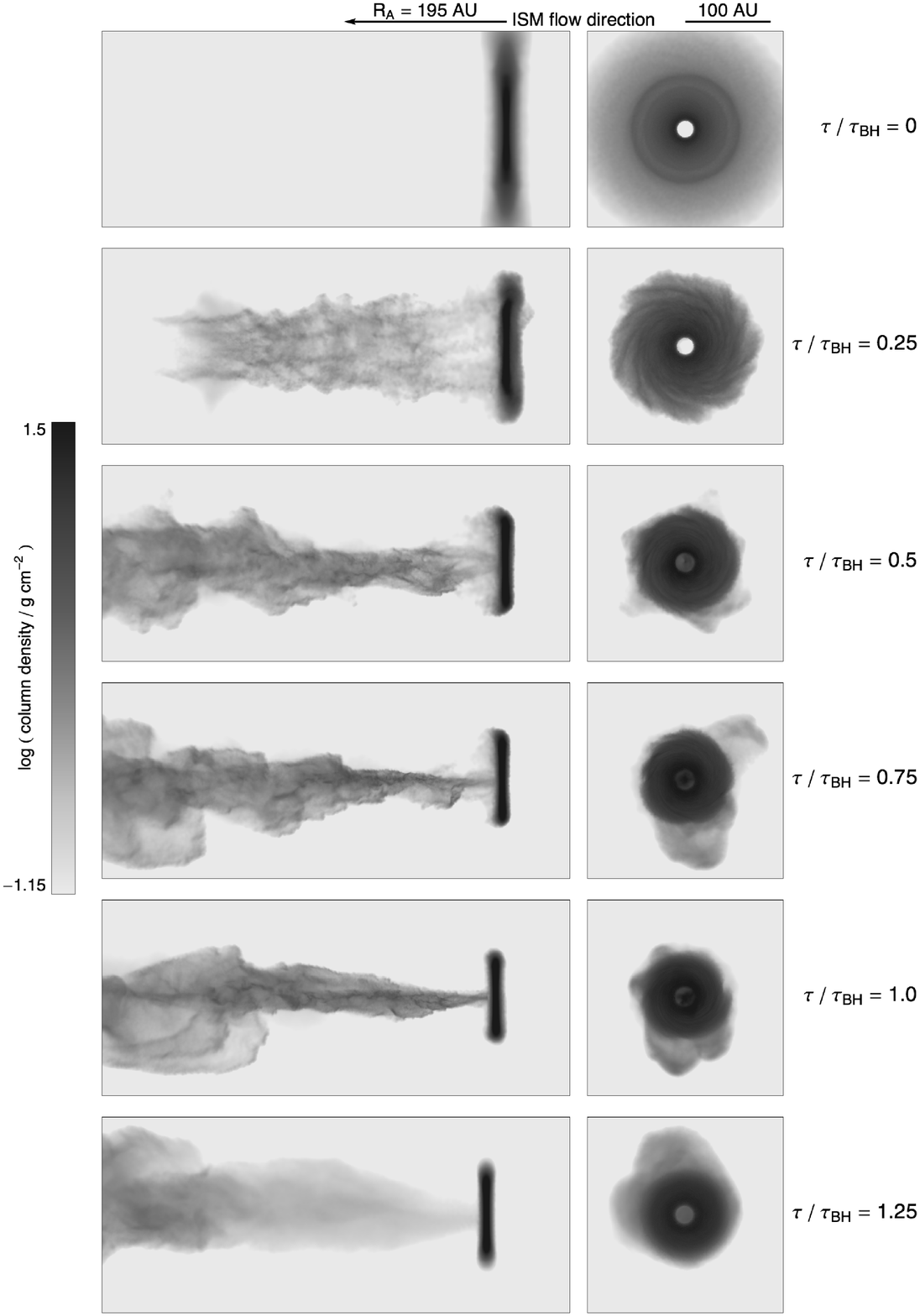}
  \caption{Snapshots from the simulation with the disk face on to the inflow, showing the column density on a logarithmic scale.  The left column is a projection orthogonal to the ISM flow vector, which is from right to left.  The right column is the projection along the direction of the ISM flow.  Time increases downward, as labeled.   Only material denser than 2.5 times the background density contributes to the column density.  An animation of this simulation can be found at {\tt http://origins.colorado.edu/{\textasciitilde}moeckel/BHLpaper/}.}
 \label{flowcomp0}
\end{figure*}
We turn now to the runs including both a disk and inflow, and consider the case of a disk with angular momentum parallel to the motion of the star through the ISM.  With this arrangement the disk is maximally absorbing the column of radius $R_A$ that the star is drilling through the ISM.  Recall that our choice of parameters has $R_{disk} / R_A \sim 0.5$.  This means that material with impact parameters $\lesssim 0.5 R_A$ will be intercepted by the disk rather than proceeding to collide downstream and accrete up the wake.  The linear momentum of this material is absorbed by the disk, leading to changes in the disk density profile and stripping of the outer regions.

Figure \ref{flowcomp0} shows the morphology of the disk and the accretion column as it develops.  We plot the projected surface density through the simulation domain on a logarithmic scale.  The top row shows the disk just before it begins interacting with the leading edge of the ISM flow.  In the second row, at time $\tau = 0.25 \tau_{BH}$, 
the accretion flow is beginning to assemble downstream of the star, and the face-on projection shows that the disk structure is changing under the influence of the accretion flow.  By $\tau = 0.5 \tau_{BH}$ the accretion column is established, and much like the pure BHL accretion case accretion proceeds up the wake and onto the accretor.  The main morphological difference is the looser collimation of the accretion column close to the star.  With the disk intercepting the inner part of the upstream flow, the wake is accreting in the lee of the disk and the strong shock from material with a small impact parameter is unavailable to constrain the flow.  This results in the wispy appearance of the accretion stream close to the star.  The stagnation point of the downstream flow, as in the diskless case, is in the vicinity of 1 $R_A$ from the star.  A slight drift in the position of the star in the direction of the ISM flow is also seen.  In our physical picture, this is due to the star slowing down as it interacts gravitationally with the wake of material behind it.

The upper right panel of figure \ref{accretionrates} shows the accretion rates onto the star from the disk material and the ISM.  The growth of the ISM accretion rate has a different shape in the presence of a disk compared to the BHL case.  In the diskless case, the early accretion as the wake is being established is due to low impact parameter material with colliding streamlines close to the star.  With a disk present, this material is intercepted and the final quasi-steady accretion flow takes longer to develop.  The low accretion rate in this case from $0 < \tau / \tau_{BH} \lesssim 0.25$ is from ISM material with impact parameter less than the accretion radius of our sink particle, which accretes from the upstream side of the star.  After $\tau \sim 0.25 \tau_{BH}$ the accretion flow generated by larger impact parameter material that avoided the disk begins to establish itself, and the ramp-up to a quasi-steady state begins.  From $\tau = 0.5 \tau_{BH}$ until the ISM flow stops the accretion rate oscillates unstably around a mean value of $2.25 \times 10^{-6}$ \msunyr (just under the reference accretion rate of equation \ref{referencerates}), though there does appear to be an increasing trend over this interval.

Accretion from the disk is largely the same as the baseline rate, with some notable spikes that coincide with peaks in the ISM accretion rate.  These episodes of increased accretion occur when a dense stream of ISM material, unconstrained radially as it approaches the star, impacts the disk at a radius of $\sim 20$ AU.  The addition of this low angular momentum material to the disk causes an accretion event.  Note that the timing and details of this process are dependent on the numerical accretion radius.  It is possible that this material would have adjusted to a smaller radius inside of our 10 AU accretion radius, and likewise this same effect (ISM accretion streams striking the disk) could occur at smaller radii and scales than we follow in these simulations.

After the flow is turned off, the remnants of the accretion stream either escape from the system or accrete onto the star, leaving the disk to equilibrate.  In the accretion rate plotted in figure \ref{accretionrates}, the accretion of the final lingering ISM wisps is seen as a non-zero ISM accretion rate, but this is not affecting the remaining disk in any significant way.  In the final panel of figure \ref{flowcomp0}, this low-density ISM material is visible in the edge-on view, with the denser material on the left edge moving away from the star toward the outflow boundary.  The face-on view shows no remaining spiral structure, as the disk has settled down to its final state.

The middle panel of figure \ref{surfacedensities} shows the surface density of the disk during this run.  Because the disk is changing shape and perhaps orientation due to the accretion flow, we define the disk for this plot as follows: considering only material within 120 AU of the star, the angular momentum of all remaining disk particles are summed.  Then all ISM particles with angular momentum orientations within $\pi/8$ of this vector are found, and included in the surface density calculation.  This procedure is effective in removing the accretion stream and isolating what one would visually deem part of the disk.  The solid lines in the plot show the total disk surface density, i.e. both the ISM and the original disk particles.  The dashed lines show only the contribution to the disk from the absorbed ISM.  

The two most obvious features of the disk evolution are the inward migration of the outer edge, and the upward trend in the surface density profile.  While it is initially tempting to attribute these effects to truncation and addition of swept up material, respectively, the dominant cause of both is the redistribution of disk material, with a contribution to the increased surface density from adding ISM material to the disk.  To illustrate this, compare the total surface density profile at the initial time to the profile at $\tau = 0.5 \tau_{BH}$ in figure \ref{surfacedensities}.  At 50 AU, the total disk surface density has increased from $\sim$3 g cm$^{-2}$ to $\sim$7 g cm$^{-2}$, while the contribution from the ISM is just under 0.9 g cm$^{-2}$; the remaining increase is from material that was originally part of the disk, but moved inwards from larger radii.  From $\tau$=1.0--1.5$\tau_{BH}$ the ISM contribution appears to decrease, but this just matches the viscous draining and spreading of the disk, as the inflow is no longer accreting onto it.  The argument against truncation being the cause of the smaller outer radius is bolstered by the amount of original disk mass remaining.  If the disk were truncated at $\sim 60$ AU, as might be guessed when looking at the plot, the remaining disk mass should be 67\% of the initial amount.  Instead, over 90\% remains, and the missing 10\% is due to accretion rather than loss to the ISM.  As estimated in section \ref{estimations}, ram pressure is ineffectual at actually stripping the disk, and inward migration of the disk due to mass loading is occurring, affecting the outer radii more than the inner disk.  The final disk has a shallower and denser profile than it started with, and depending on the radius $\sim$10--25\% of the disk material came from the ISM.

The full situation is of course more complicated than our simple estimates suggested.  While ram pressure cannot remove the outer radii of the disk, it does have an early influence on the disk morphology, as is clearly seen when viewing a movie of the simulation snapshots \citep[and perhaps in the real universe, as in HD 61005;][]{hines07}.  The accretion flow has a velocity component that is directed radially inward by the time it hits the disk.  The outermost 10 AU or so to remain bound to the star-disk system, but they are deflected along the flow in toward the star, rejoining the disk plane inward of their original location and hastening the radial evolution compared to our estimated rate.

\subsection{BHL Accretion With Disk Inclined $30^{\circ}$ To Inflow}
\begin{figure*}
\centering
  \epsscale{0.75}
  \plotone{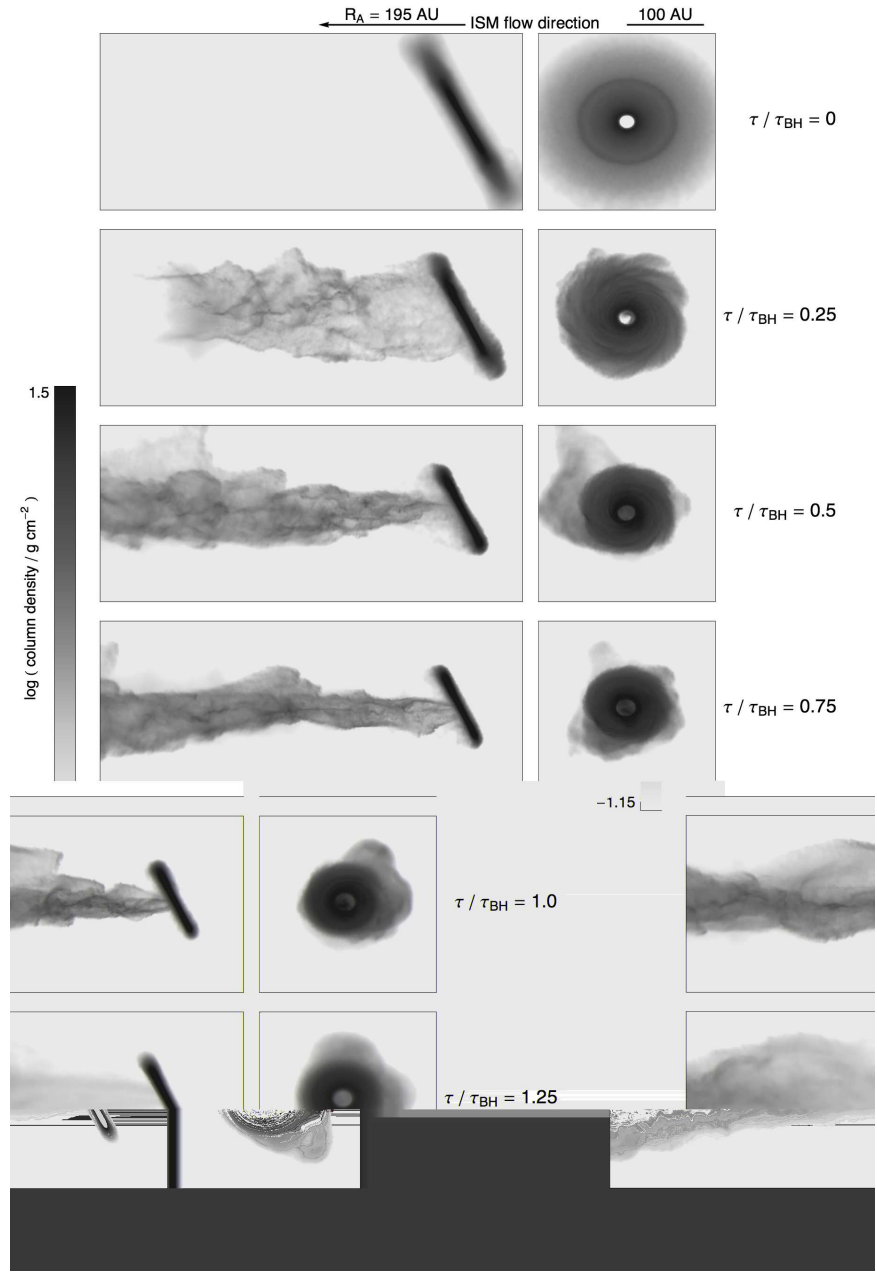}
  \caption{Snapshots from the simulation with the disk inclined $30^\circ$ to the inflow, showing the column density on a logarithmic scale.  The left column is a projection orthogonal to the ISM flow vector, which is from right to left.  The right column is the projection along the direction of the ISM flow.  Time increases downward, as labeled.   Only material denser than 2.5 times the background density contributes to the column density.  An animation of this simulation can be found at {\tt http://origins.colorado.edu/{\textasciitilde}moeckel/BHLpaper/}.}
 \label{flowcomp30}
\end{figure*}

At a bulk level, the accretion flow in the inclined case (shown in figure \ref{flowcomp30}) is very similar to the face-on case, with a similarly shaped accretion column leading up to the star, and a stagnation point as expected near $R_A$ downstream of the star.  The similarity is expected, since except for some minor geometrical effects from the disk inclination the situations are nearly identical, with the inner half of the accretion radius intercepted by the disk.  The accretion rates from the disk and the ISM, plotted in the lower right panel of figure \ref{accretionrates}, is also quite similar, with nearly identical average accretion rates from $\tau$=0.5--1$\tau_{BH}$.  Spikes in the ISM accretion rate are matched by spikes in the disk accretion rate, as the inner disk absorbs the low specific angular momentum material of the accretion column and is accreted.  The same caveats about the size of our accretion radius apply to this case as the face-on simulation.

While the accretion flow is largely the same between the two cases, one may expect the disk evolution to be different.  The component of the Keplerian disk velocity in the direction of the inflow is greater than $\sim 2$ \kmss throughout the disk, and indeed greater than the inflow velocity of 3 \kmss inward of $\sim 25$ AU.  While the absorption of the smooth inflow from the upstream direction should be averaged over the orbital period and negate this effect of the inclination, the effect of absorbing the dense and filamentary accretion column may depend on where it strikes the disk in a way not seen in the face-on case.  Examining the surface density evolution in figure \ref{surfacedensities} it appears that there may be some difference in the later evolution of the inner disk, but the difference is small enough that run-to-run variability could be the main cause.  Future investigations will pursue this matter further.

\section{Discussion and Conclusions}
As discussed above, we were constrained by computational necessity to consider a disk interacting with a dense molecular ISM, corresponding perhaps to a filament or core in a star-forming region.  The analysis of TB08 which motivated this work dealt with an ISM density reflecting the mean value of a region, much lower than this work ($10^3$--$10^6$ cm$^{-3}$), and over longer timescales (Myr versus kyr).  Because of the nature of the interactions, we believe that the main results of this work should carry over from the dense interactions we simulated to the lower density, longer timescale interactions from TB08.  If ram pressure had been the dominant driver of the changes to the disk morphology, this would not be the case, so interactions with a much denser region or at much higher velocities will be qualitatively different than these.  However, the changes were instead chiefly a result of the deposition of material with no azimuthal angular momentum onto the disk over timescales longer than (or comparable to) the orbital timescale.  These effects should remain broadly the same, whether the low specific angular momentum material is incorporated into the disk over 10 or 1000 orbital periods.  Likewise, if the angular momentum is deposited episodically (as in TB08) versus in one interaction as we have modeled, the final effect should mostly be a function of the total mass deposited. Thus the main result of this work, the redistribution of the outer disk material inward, should be robust over a wide range of ISM parameters.

A potentially significant complication that we did not consider is that the ISM in a real star-forming cloud will not be smooth, but instead filamentary and clumpy.  To some extent we have modeled this, as the star moved from a region of zero density into the dense region, and left after moving through it for $\sim 3000$ yr (equivalent to a core or filament $\sim0.01$ pc across).  However, the effect of substructure on smaller scales would be an interesting direction for future studies.  Density or velocity structure on scales comparable to the accretion radius could have an important effect on how the disk restructuring proceeds, introducing net angular momentum to the inflow and perhaps changing the disk structure more dramatically than the almost totally-smooth ISM we have considered.  Simulations of density and velocity gradients in pure BHL accretion \citep{ruffert97,ruffert99} show that strong angular momentum gradients can result in small, short-lived disk-like structures as the flow accretes onto the star.  The interaction of such an accretion flow with an extant disk warrants further investigation.

Another issue that is unmodeled here is the potential effect of protostellar outflows on the accretion flow.  There are two main effects which could alter the flow; radiation pressure, and mechanical winds or jets.  The influence of stellar radiation on BHL accretion has been investigated by \citet{edgar04a}, who showed that for stellar masses above around 10 \msuns the accretion rate can be significantly diminished, but for lower mass stars like those in this paper radiative feedback should not present a large impediment to accretion.  The question of jets' and winds' effects on an accretion flow was briefly considered in TB08.  The ram pressure of a jet from a typical T Tauri star could be enough to reverse the BHL flow.  However, the opening angle of collimated outflows tends to be $\sim20^\circ$ at a distance of 20-50 AU, and smaller at larger radii \citep{hartigan04}.  The solid angle directly affected by a jet is then less than about 1 sr at scales comparable to the accretion radius, and assuming that the flow intersecting the jet is reversed, this effect is on the order of 10\%.  The zero-inclination case we simulated would be more greatly affected, as the jet would be directly opposing the accretion wake.  This would most directly alter the accretion rate onto the star, however, and the accretion onto the disk would be less impacted.  To our knowledge the interaction between a collimated outflow and a BHL accretion flow has not been modeled numerically, and this presents and interesting question for further work, but we estimate that it should be a secondary effect.

With these thoughts about the broader applicability of our models in mind, we discuss the main results of our work and their implications for disk evolution and planet formation: 

{\em Changes to the disk surface density profile and morphology}.  We have identified two effects that change the surface density profile of the disk: accretion onto the disk from the ISM flow, and the subsequent redistribution of the extant disk material due to absorbing low angular momentum material.  While related in that the accretion of ISM material leads directly to the redistribution of the existing disk, the latter effect is more important in terms of changes to the disk surface density profile.  With accretion alone the surface density profile would increase $\sim$10--20\% (see figure \ref{surfacedensities}), while with the migration of outer material inwards the profile more than doubles across much of the disk relative to the initial value.

A possible consequence of this, pointed out in TB08, is the delivery of fresh material to regions of the disk that may have been depleted by planet formation, possibly ushering in a second era of gas accretion.  TB08 note that over several Myr accretion can deliver an amount of gas comparable to (or even greater than) the original disk mass.  This work suggests that the effect of accretion could be magnified by the presence of remnant gas disk, dumping not only the accreted ISM gas but the outer disk gas onto the inner disk and any planets that may be forming there.  The presence of a dissipative disk can have an important influence on the dynamics of multi-planet systems \citep{moorhead08,moeckel08,thommes08,matsumura09}, and the dynamics of such systems in a disk that is undergoing induced global migration may prove interesting.

A secondary effect is the shrinking outer radius of the disk.  As long as the disk is accreting ISM material, the inward migration of the outer regions continues, as is seen in figure \ref{surfacedensities}.  Any wispy outer portions of the disk are more susceptible to ram pressure, and thus a clean disk edge is maintained.  This effect would be most pronounced in disks that have undergone more accretion, perhaps due to being in a denser-on-average region, for instance the center of a cluster potential.  As the disk redistribution occurs on a timescale comparable to the orbital period, this process could take effect quickly and perhaps provides a mechanism to generate a relationship between disk-radius and the system's radius in a cluster.  This would work in the same sense as later photoevaporation once a cluster's O stars begin to dominate their locality, for example as in NGC2244 \citep{balog07}.

\begin{figure}
\centering
  \plotone{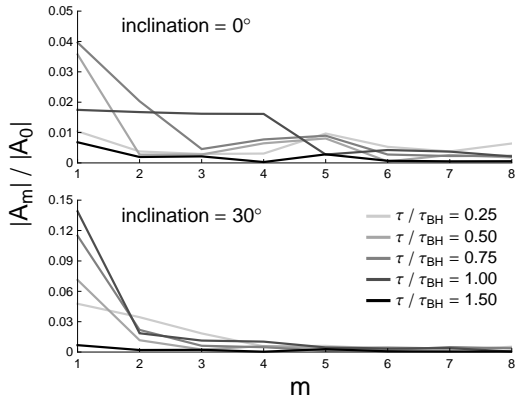}
  \caption{Amplitudes of the first eight Fourier components in the azimuth of the disk surface density.  {\it Top}: the face-on case, showing some power in modes with $m\lesssim5$, dominated by $m=1$.  {\it Bottom}: the $30^\circ$ inclination case, with a stronger $m=1$ mode due to the misalignment between the inflow vector and disk symmetry axis.  Note the difference in scale between the two plots.}
  \label{fourier}
\end{figure}
{\em Triggered gravitational instabilities?} With our chosen parameters, the disks should be very stable against gravitational instabilities that may aid planet formation \citep[e.g.][]{boss97,boss03,rice06}.  Despite this, we see spiral features in the column density plots (figures \ref{flowcomp0} and \ref{flowcomp30}).  To better understand these, we examine the azimuthal Fourier amplitudes of the surface density, which we compute as 
\begin{equation*}
  A_m = \sum_{j=1}^{n_{disk}} m_{j}  e^{-i m \phi_{j}} ,
\end{equation*}
where $m_j$ is the mass of the SPH particle, $m$ is the mode, $\phi_{j}$ is the azimuthal angle of the particle, and the summation is over the particles identified by their angular momentum as part of the disk.  In figure \ref{fourier} we plot the amplitude of the first eight Fourier modes (normalized by $A_0$) for each of the runs, at five times each.  The Fourier modes for the isolated disk, which we do not plot, lack any power above $|A_m|/|A_0| = 0.003$ in the first eight modes, confirming that absent the influence of the ISM flow, these disks are smooth and stable.  In the face-on case, the spiral structure begins with a marginal increase in the amplitude of the $m \sim$ 4--5 modes which are tracing the spiral features most clearly seen in the top panel of figure \ref{flowcomp0}.  The low amplitudes suggest that these spiral features are not globally organized features due to gravitational instability, but rather due to a more local reorganization of the disk as the outer material is redistributed inwards.  

The more striking feature is the $m=1$ mode, which dominates the Fourier spectrum at later times for $i=0^\circ$ and at all times for $i=30^\circ$.  This is due to the star accreting a filament of material with some momentum perpendicular to the accretion axis, so that the star is knocked slightly out of the center of the disk.  As the disk responds to this off-axis potential, the $m=1$ mode is excited.  The signal is much stronger in the inclined case, as even accretion of material along the inflow direction will move the star slightly out of the disk's axis of symmetry.  Thus this signal, too, is not due to gravitational instability, and the spiral features are unlikely to be signals of disk fragmentation leading to planet formation.  We note, however, that a disk beginning in a more marginal state of stability than ours could be pushed into an unstable state by the disk redistribution we see.  If a factor of $\sim$2 increase in surface density would be enough to push a disk into an unstable state, than interactions with the ISM like we have modeled here may be sufficient to stimulate planet formation by gravitational collapse.  For the low-mass disk simulated here though, we stress that even with the addition of mass from the ISM the Toomre Q parameter remains much larger than unity.  The spiral features are {\em not} due to gravitational instability, but instead the interaction of the disk and the star with an accretion flow.  This process could conceivably masquerade as gravitational instability in observations.

{\em Changes to the disk inclination}.
Intuitively the addition of material with angular momentum misaligned with the rotation of the disk seems like a potential route toward the inclination difference between the ecliptic plane and the Sun's rotation axis.  The instantaneous orientation of the disk, measured by the average angular momentum of the particles identified as belonging to the disk as in the surface density calculation, is a bit noisy, but at the end of the simulations when the disk has settled down the face-on case exhibits a $0.5^\circ$ inclination change relative to its initial orientation, and the inclined case has changed by $1.5^\circ$.  This mechanism would appear to not be enough to account for the Sun's $\sim7^\circ$ inclination relative to the ecliptic plane \citep{beck05}, at least with the amount of mass accreted in our simulations.  If the accreted ISM material overwhelms the initial disk, as shown to be possible in TB08, the situation might be different, but a much more detailed examination of the timescales for planet formation versus disk replacement and reorientation would be necessary to make this claim.

To conclude, we have performed simulations studying the interaction between a protoplanetary disk and  a dense, ambient ISM.  The accretion rate from the ISM onto the star is fairly well described by the standard BHL accretion rate, even in the presence of a disk altering the flow.  We have showed that these interactions can have a significant effect on the physical structure of the disk, with the main effect being the redistribution the outer disk inwards.  This redistribution, and the addition of foreign material to the extant disk, may have some implications for planet formation scenarios.\\  

\acknowledgements
This research was partially supported by the National Science Foundation through TeraGrid resources provided by NCSA.  NM thanks Ian Bonnell for helpful discussions, and the administrations at the Universities of St Andrews and Exeter for arriving at the groundbreaking agreement that facilitated his joint appointment during this work.  HT acknowledges support from NASA Origins grants NNG05GI43G and NNG06GH33G.

\end{document}